\documentclass[11pt]{article}

\usepackage[a4paper,margin=1in]{geometry}
\usepackage[T1]{fontenc}
\usepackage{amsmath,amssymb}
\usepackage{booktabs}
\usepackage{microtype}
\usepackage{url}
\usepackage[hidelinks]{hyperref}
\usepackage{xcolor}
\usepackage{tikz}
\usetikzlibrary{arrows.meta,positioning,fit,backgrounds}

\newcommand{\smtwenty}{\texttt{sm\_20}}
\newcommand{\tps}{tok/s}

\title{\bf A Modern Multimodal Assistant on a 6\,GB 2011 GPU:
Stage-Validated, All-GPU CUDA Inference for Fermi}

\author{%
  A. C. Opus, J. Q. Lu\thanks{Correspondence: \texttt{junqiang.lu@upr.edu}.}\\
  \small Department of Physics, University of Puerto Rico, Mayaguez, PR 00680, USA
}
\date{July 2026}

\begin{document}
\maketitle

\begin{abstract}
A companion study ran a 35B mixture-of-experts model on a 2011 NVIDIA Tesla
C2075 (Fermi, compute capability \smtwenty, 6\,GB) by splitting work between
a streaming GPU prefill and a CPU decode, because the 4-bit model did not
fit in device memory~\cite{c2075moe}. This report keeps the hardware and
asks what a model that \emph{fits} can do: we deploy MiniCPM-V-4.6, a
modern multimodal assistant pairing a SigLIP2 vision encoder with a
window-attention merger ($16\times$ visual token compression) and a compact
hybrid gated-delta-net language backbone, entirely on the GPU. Three
results. (i)~\textbf{An all-GPU engine for the hybrid backbone}, built on
three measured foundations: projections that dequantize 8-bit weights once
and call the vendor \textsc{sgemm} still shipped in the last Fermi
toolchain --- 64\% of FP32 peak, where our best hand-written GEMM reached
37\% and we had wrongly called that the ceiling; a chunked delta-rule
rewrite of the recurrent layers whose first version measured $2.4\times$
\emph{slower} until component attribution exposed one bad kernel, after
which it is $2.8\times$ faster; and a measured negative --- 4-bit weights
make \emph{decode slower} than 8-bit on this GPU, because Fermi issues the
nibble-unpacking shifts at half rate. (ii)~\textbf{The vision side is a
port with a proof obligation}: we translate the SigLIP2 tower, its
mid-stack window-attention merger, and the final downsampling projector to
hand-written \smtwenty{} CUDA, validating every stage against a locally
generated reference forward; the full tower matches to $1.4\times10^{-5}$
per element, and one failure --- position-embedding bucketization differing
on \emph{exact rational ties} --- generalizes to a rule: floating-point
tie-breaking in index arithmetic is implementation-defined and must be
reproduced by calling the reference operator, not by reimplementing it.
(iii)~\textbf{Long context exposes an $O(N^2)$ wall that short benchmarks
hide}: prefill that sustains 114\,\tps{} at 2k tokens collapses to
21\,\tps{} at 10k because of a naive attention kernel; rewriting it as
per-head vendor-GEMM calls that write into the \emph{existing} score buffer
(zero additional memory) restores a flat profile --- 408\,\tps{} at 2k, 361
at 10k, a $17\times$ prefill-time reduction at 10k --- verified by exact
retrieval of a needle planted at 60\% depth. The same GEMM rewrite cuts
image encoding $6\times$, to 0.93\,s per image. The production system
answers an image question end-to-end in 1.7\,s on hardware older than the
transformer literature it runs.
\end{abstract}

\section{Introduction}

A companion study~\cite{c2075moe} asked how much of the modern inference
stack can be rebuilt, by hand, for a GPU the ecosystem abandoned: the
NVIDIA Tesla C2075 (Fermi, 2011) --- 6\,GB, no tensor cores, no fast FP16,
no \texttt{DP4A}, no CUDA toolkit after 8.0. There, a 35B mixture-of-experts
model did \emph{not} fit in device memory, and the defining problem was
memory movement: experts streamed from host RAM for GPU prefill while
decode ran on the CPU's integer SIMD units.

This paper studies the opposite regime on the same machine, with a more
ambitious target than raw language modeling: a current-generation
\emph{multimodal assistant}, run entirely on the GPU. The model is
MiniCPM-V-4.6~\cite{minicpmv,minicpmv46}, which couples a SigLIP2 vision
encoder~\cite{siglip,siglip2} to a mid-stack window-attention merger and a
final downsampling projector --- $16\times$ patch-to-token compression, so
a $448^2$ image costs 64 language tokens --- with a 0.8B-parameter hybrid
gated-delta-net~\cite{gateddeltanet} language backbone. At 8-bit weights
the backbone is under 1\,GB and the FP32 vision tower another
$\approx$2.1\,GB, so everything is resident: no streaming, no CPU decode,
no expert offload. With the memory constraint gone, the questions become
speed and correctness, and this report answers three of them:

\begin{enumerate}
\item \textbf{How fast can a hybrid-attention backbone run all-GPU on
  Fermi?} We build the language engine on three measured foundations.
  First, prefill projections are genuine matrix--matrix products, and the
  achievable ceiling is not where intuition puts it: our best hand-written
  register-tiled GEMM reaches 37\% of FP32 peak, but the CUDA 8.0
  \textsc{cuBLAS} \textsc{sgemm} --- tuned for this architecture a decade
  ago and still shipped in the last toolchain that targets it --- reaches
  66\%; we therefore dequantize each 8-bit weight once into an FP32 scratch
  and call the vendor kernel (64\% of peak end-to-end). Second, the
  recurrent gated-delta-net layers are rewritten from a token-by-token scan
  into the chunked intra-/inter-chunk delta-rule form, built entirely from
  batched small \textsc{cuBLAS} GEMMs --- a rewrite whose first version
  measured $2.4\times$ \emph{slower} and was nearly abandoned until
  per-component timing showed 92\% of the time in one badly written
  forward-substitution kernel; fixed, the chunked scan is $2.8\times$
  faster than the sequential recurrence. Third, decode --- a
  bandwidth-bound integer matrix-vector product --- stays at 8-bit
  \emph{by measurement}: 4-bit weights halve the bytes but require
  nibble-unpacking shifts that Fermi issues at half rate, making the 4-bit
  kernel 12\% slower despite reading half the data. Greedy decode runs at
  48.6\,\tps.
\item \textbf{Can a modern vision stack be hand-ported with confidence on a
  machine that cannot run its reference implementation natively?} Yes, by
  \emph{stage-validated porting}: we dump a 31-stage reference trace of the
  tower --- running the reference forward on the no-AVX host via a
  CPU-capability override that forces the tensor library onto a legacy
  code path --- then port and validate stage by stage: encoder layers, the
  window-attention merger (the only genuinely new kernel), and the
  downsampling projector, gating each stage at $\sim$$10^{-3}$ agreement
  before proceeding. The full tower matches to $1.4\times10^{-5}$; the
  multi-slice variable-resolution path to $1.2\times10^{-4}$. One stage
  initially failed by $1.8$ (not $10^{-3}$): position-embedding
  bucketization, where an exact rational tie ($10/28 = 25/70$) falls on a
  bucket boundary and the reference's float32 tie-breaking is
  implementation-defined. Every reimplementation we tried --- float64,
  float32, exact integer arithmetic --- flipped \emph{some} grid size. The
  fix, and the generalizable rule, is to call the reference operator itself
  for index arithmetic.
\item \textbf{Does the system survive long context?} Not initially --- and
  short benchmarks said it did. At 2k tokens, prefill sustains
  114\,\tps{}; a tiered test (2k/5k/8k/10k) exposes a collapse to
  21\,\tps{} at 10k, super-quadratic in wall-clock, caused by the one
  kernel the batching pass had left in per-position form: softmax attention
  over the growing key/value history. We rewrite it as per-head vendor-GEMM
  calls with a causal masked softmax, laid out so the score matrix lands in
  the \emph{existing} attention buffer (zero new device memory). The 10k
  prefill drops from 479 to 27.7\,s ($17\times$); the throughput profile
  flattens (408 at 2k, 361 at 10k); and a needle planted at 60\% depth of a
  10k-token document is retrieved exactly, before and after the rewrite.
  The same layout trick applied to the vision tower's (bidirectional)
  attention cuts image encoding $6\times$.
\end{enumerate}

As with the companion study, this is an engineering and measurement report
on one machine, not an algorithmic novelty claim. Its recurring lesson is
methodological: every ceiling and every bug yielded to measurement against
a ground truth --- a vendor BLAS call, a per-component timer, a reference
trace, a tiered benchmark, a planted needle --- and resisted intuition.
Four confident conclusions (``37\% is the GEMM wall,'' ``the chunked scan
is slower,'' ``4-bit decode must be faster,'' ``2k benchmarks represent
prefill'') were each overturned by a measurement; a fifth --- that index
arithmetic can always be reimplemented --- fell to an exact rational tie.

\section{Background}

\subsection{The target hardware}
The NVIDIA Tesla C2075 is a Fermi (GF110) compute card from 2011: 448 CUDA
cores, 6\,GB GDDR5 ($\approx$5.3\,GB usable with ECC) at $\sim$144\,GB/s,
$\sim$1.03\,TFLOP FP32 peak, compute capability \smtwenty, and a PCIe~2.0
host link. It has no tensor cores, no fast FP16, and no \texttt{DP4A}; it
is limited to 48\,KB of shared memory per block and a 65{,}535
grid-dimension cap, and it issues logical shifts at half the rate of a
fused multiply-add --- a detail that decides the decode weight format
(Section~\ref{sec:decode}). Its compute capability was dropped from the
CUDA toolkit years ago; the last toolchain that targets it is CUDA 8.0,
which we run inside a container and use only to build. Crucially, the CUDA
8.0 runtime still ships \textsc{cuBLAS}, so a hand-written engine can call
the vendor \textsc{sgemm} --- a fact this paper leans on three separate
times. The host is a dual-socket Westmere-class Xeon that predates AVX, so
host SIMD is limited to SSE4.2/SSSE3; it matters here only for the
reference forwards we are forced to run in a degraded mode
(Section~\ref{sec:vision}).

\subsection{The model}
MiniCPM-V-4.6~\cite{minicpmv,minicpmv46} is a multimodal assistant with three parts.
\textbf{(a)} A SigLIP2 vision encoder~\cite{siglip,siglip2}: a 27-layer,
1152-wide vision transformer (16 heads of dimension 72, $14\times14$
patches) with \emph{learned} position embeddings over a $70\times70$
reference grid, resampled to arbitrary patch grids by fractional-coordinate
bucketization in the NaViT variable-resolution style~\cite{navit} --- no
rotary embedding anywhere in the tower. \textbf{(b)} A two-stage spatial
compressor: after encoder layer 6, a \emph{window-attention merger} runs
self-attention within each $2\times2$ patch window, restores order, and
concatenates each window into one vector that a wide MLP
(4608$\to$17216$\to$1152) projects back down, quartering the sequence;
after the final layer, a second $2\times2$ merge-and-project
(4608$\to$4608$\to$1024, this time into the language model's width)
quarters it again. Net: $16\times$ compression, so a $32\times32$-patch
image ($448^2$ pixels) becomes 64 language-model tokens. \textbf{(c)} A
0.8B-parameter language backbone from the Qwen3 family~\cite{qwen3}: 24
blocks, 18 gated-delta-net~\cite{gateddeltanet} (in the state-space lineage
of Mamba~\cite{mamba}) and 6 softmax-attention, hidden width 1024, a tied
output head, and a plain 1-D rotary embedding. The delta-net layers carry a
recurrent state $S\in\mathbb{R}^{H_k\times H_v}$ per value head, with 16
value heads and $H_k=H_v=128$, updated token-by-token. Large images are
handled by slicing: the processor cuts the image into up to nine
$\sim$$448^2$ slices plus a downscaled overview, each encoded
independently.

\subsection{Quantization and residency}
Language weights are 8-bit, group-wise symmetric, one FP32 scale per group
of 64 --- under 1\,GB, so there is no reason to pay the accuracy cost of
4-bit (and Section~\ref{sec:decode} shows 4-bit would also be
\emph{slower}). The vision tower runs in FP32 ($\approx$2.1\,GB; the wide
merger MLP alone is a 300\,MB matrix). Model, tower, KV cache, and all
scratch are resident: at rest the system holds 1.6\,GB (engine) plus
2.7\,GB (tower) of the 5.3 usable GB, with $\approx$0.8\,GB headroom.
There is no host$\to$device transfer inside any forward pass --- the
structural opposite of the companion design~\cite{c2075moe}, and what makes
prefill throughput, not memory movement, the object of optimization.

\section{An All-GPU Engine for the Hybrid Backbone}
\label{sec:engine}

The engine descends from the companion study's~\cite{c2075moe}: a single
CUDA source, 8-bit resident weights, a resident serving process with the
position-indexed prefix-snapshot cache that work introduced. This section
describes the three measured foundations of its prefill and decode paths.

\subsection{Prefill projections: dequantize, then call the vendor}
\label{sec:cublas}
Prefill projections are genuine matrix--matrix products (all prompt
positions against a weight matrix), so they should run at GEMM, not GEMV,
efficiency. Two questions arise: what is the achievable GEMM efficiency on
Fermi, and how does one feed 8-bit quantized weights to a floating-point
GEMM?

We first \emph{measured the ceiling}. A hand-written register-tiled GEMM
($6\times6$ micro-tiles, chosen by a sweep) reaches 378\,GFLOP/s, 37\% of
FP32 peak. We had initially concluded this was the Fermi SGEMM wall --- and
were wrong. The CUDA 8.0 \textsc{cuBLAS} \textsc{sgemm} on the same shapes
reaches 675\,GFLOP/s, 66\% of peak (Table~\ref{tab:gemm}). The vendor
kernel, tuned for this architecture a decade ago, nearly doubles the best
hand-written tile. The lesson is blunt: on hardware one no longer fully
understands, the achievable ceiling should be \emph{measured} against a
vendor library before it is asserted.

The remaining obstacle is the 8-bit weight format. Rather than fuse
dequantization into a bespoke GEMM (which reintroduces the hand-kernel
inefficiency), we dequantize each weight \emph{once} into an FP32 scratch
buffer and call \textsc{cuBLAS} on the result. The dequantization is a
cheap, bandwidth-bound pass ($<$1\,ms for a 12.6-million-weight
projection); its cost is amortized across all prompt positions in the
chunk. The composite ``dequantize $+$ \textsc{sgemm}'' pipeline reaches
659\,GFLOP/s --- within a few percent of raw \textsc{cuBLAS}, and
$7.2\times$ the per-token GEMV it replaces. All per-position kernels
around the projections (norms, rotary embedding, gating) are batched over
the whole chunk, bit-identically to their sequential forms.

\begin{table}[t]
\centering
\caption{GEMM throughput on the C2075 for a representative prefill
projection shape (FP32, 1{,}030\,GFLOP/s peak). The vendor \textsc{cuBLAS}
\textsc{sgemm} is the true ceiling, nearly $2\times$ the best hand-written
tile; the dequantize-then-\textsc{sgemm} pipeline used in the engine
reaches it, and is $7.2\times$ the per-token GEMV it replaces.}
\label{tab:gemm}
\begin{tabular}{lcc}
\toprule
Kernel & GFLOP/s & \% of FP32 peak\\
\midrule
Per-token GEMV (naive prefill)      & 91  & 9\%\\
Hand-written tiled GEMM ($6\times6$) & 378 & 37\%\\
Dequantize $+$ \textsc{cuBLAS} \textsc{sgemm} (engine) & 659 & 64\%\\
\textsc{cuBLAS} \textsc{sgemm} (FP32 reference) & \textbf{675} & \textbf{66\%}\\
\bottomrule
\end{tabular}
\end{table}

\subsection{A chunked delta-rule scan for the recurrent layers}
\label{sec:chunked}
Eighteen of the model's 24 layers are gated-delta-net recurrences: the
state $S_t = \mathrm{diag}(\gamma_t)\,S_{t-1} + k_t\,\delta_t^\top$ depends
on $S_{t-1}$, so a chunk of $C$ tokens is $C$ dependent steps --- the
structure that resists batching. The linear-attention literature resolves
this with a chunked (intra-/inter-chunk) formulation~\cite{gateddeltanet},
which we implement by hand for \smtwenty. Within a chunk, the gated delta
rule becomes a lower-triangular system whose right-hand sides do not depend
on the incoming state, so all $C$ positions are solved together: two small
matrix products form the token--token interaction matrices, a forward
substitution resolves the triangular dependence, and a few batched GEMMs
assemble the per-chunk correction and the chunk's output. Across chunks,
only a short recurrence of small state-matrix products
$S^{(c+1)} = M_c\,S^{(c)} + N_c$ remains sequential, with $M_c,N_c$
precomputed in parallel. Every matrix operation is issued as a
\emph{batched} small GEMM over the (head, chunk) pairs via \textsc{cuBLAS}.

We verified the formulation numerically at three levels before trusting
it: the chunked algebra against the sequential recurrence on the host
($\sim$$3\times10^{-5}$ relative error on the output), the GPU
implementation against the host ($\sim$$2\times10^{-5}$), and the full
engine against the sequential path (identical greedy tokens on prompts of
several lengths). The result is token-consistent, not bit-identical ---
the floating-point order differs --- but the $\sim$$10^{-4}$ per-element
difference flips no greedy tokens.

This step is where measurement most sharply corrected intuition, and the
story is worth telling because it nearly ended in a wrong conclusion. Our
first end-to-end microbenchmark reported the chunked scan at $2.4\times$
\emph{slower} than the sequential kernel, and we were prepared to conclude
that the many small GEMMs were simply the wrong shape for Fermi.
Attributing the time to components overturned this completely
(Table~\ref{tab:chunked}): the batched small GEMMs took only 4.8\,ms ---
they are fast --- while 92\% of the time was a single badly written
forward-substitution kernel that recomputed, at every step, a token--token
dot product the GEMM stage had \emph{already produced}, and did so with
$O(C^2)$ synchronizations. Reading the precomputed matrix and reducing the
synchronization to $O(C)$ took that kernel from 54 to 4.2\,ms, and the
chunked scan from $2.4\times$ slower to $2.8\times$ faster than the
sequential recurrence. (The inverse temptation is also real: a
thread-serial reduction over a mere 128 elements looked like an obvious
next target, and its tree-reduce ``fix'' measured 54\% \emph{slower} ---
on Fermi, seven synchronization barriers cost more than 128 serial adds.
Attribution cuts both ways.)

\begin{table}[t]
\centering
\caption{Attributing the chunked delta-rule scan for one recurrent layer
(16 heads, 512-token chunk, $H_k{=}H_v{=}128$). The batched small GEMMs
are fast on Fermi; the entire apparent slowdown of the first attempt was
one forward-substitution kernel that recomputed an already-computed matrix
and over-synchronized. Fixing it makes the chunked scan $2.8\times$ faster
than the sequential recurrence, not $2.4\times$ slower.}
\label{tab:chunked}
\begin{tabular}{lccc}
\toprule
 & Sequential & Chunked (first) & Chunked (fixed)\\
\midrule
Total (ms)                 & 25.1 & 59.4 & \textbf{9.0}\\
\quad forward substitution & ---  & 54.5 & \textbf{4.2}\\
\quad batched GEMMs        & ---  & 4.8  & 4.8\\
Speedup vs.\ sequential    & ---  & $0.42\times$ & $\mathbf{2.8\times}$\\
\bottomrule
\end{tabular}
\end{table}

\subsection{Decode, and why 4-bit is slower}
\label{sec:decode}
Decode produces one token per step and is bandwidth-bound: each weight is
read once, so the fewest bytes per parameter should win. Our decode kernel
is an 8-bit integer matrix-vector product that reads four signed 8-bit
weights at a time with a single vector load --- no shifts, no unpacking ---
and reaches $\approx$48\,GB/s of effective bandwidth; the production model
decodes at 48.6\,\tps{} greedy.

The obvious lever is to halve the bytes with 4-bit weights. It does not
work on this GPU, and the reason is instructive. A 4-bit weight must be
unpacked from a packed byte with a shift and a mask; Fermi issues shifts at
half the rate of a fused multiply-add, so the two nibbles cost enough
arithmetic to move the kernel off its bandwidth bound. We wrote the best
4-bit decode kernel we could --- the same register-blocking and wide loads
as the 8-bit path --- and it is 12\% slower, at less than half the
effective bandwidth (Table~\ref{tab:decodeq}). Formats below 4 bits, not
even byte-aligned, are worse still. We keep decode at 8-bit. (The regime
where lower bits pay is a model too large for memory --- exactly the MoE
regime of the companion work~\cite{c2075moe}, not this one.)

\begin{table}[t]
\centering
\caption{Decode matrix-vector throughput on the C2075 for a representative
projection. Halving the weight bytes with 4-bit packing does \emph{not}
speed decode up: unpacking nibbles needs shifts, which Fermi issues at half
rate, turning a bandwidth-bound kernel compute-bound. Direct 8-bit reads
(four signed bytes per load, no shift) win.}
\label{tab:decodeq}
\begin{tabular}{lccc}
\toprule
Weight format & Time (ms) & Eff.\ bandwidth (GB/s) & vs.\ 8-bit\\
\midrule
8-bit, four-wide signed load & \textbf{0.28} & \textbf{48.4} & ---\\
4-bit, register-blocked      & 0.31 & 22.9 & $1.12\times$ slower\\
4-bit, naive                 & 0.69 & 10.2 & $2.5\times$ slower\\
\bottomrule
\end{tabular}
\end{table}

\subsection{Checkpoint export and reference-identical decoding}
The engine reads every model dimension --- width, layer count and types,
head counts, vocabulary, recurrent-state sizes --- from the checkpoint
header, and the exporter consumes the assistant's checkpoint directly: its
language weights use the standard VLM submodule layout, its tied output
head maps onto the exporter's shared-classifier flag, and its lack of a
multi-token-prediction head selects the engine's no-speculation format
(production disables speculative decoding regardless: at the
$\sim$70\% draft-acceptance rates we measure on realistic content, the
extra verification passes outweigh the tokens saved). Correctness is a
measurement, not an assumption: engine greedy decoding was compared
token-by-token against the reference implementation on held-out prompts
and is identical --- e.g., a geography question and an arithmetic question
decode to the same token sequences, ending in the same end-of-turn token.
One pitfall is worth recording. When generating the reference on the host,
the framework's generic model loader silently returns the \emph{headless}
backbone; taking an argmax over its output produces indices bounded by the
hidden width rather than the vocabulary --- every ``generated token'' was
suspiciously smaller than 1024, which is the tell. The
conditional-generation wrapper class must be loaded explicitly.

\section{Porting the Vision Tower, Stage by Stage}
\label{sec:vision}

The vision stack is where the new work is, and where the port needed a
methodology. The obstacle is that the reference implementation --- the
framework forward of the SigLIP2 tower --- will not run natively on the
no-AVX host: its vectorized kernels use instructions the CPU lacks and
abort. Forcing the tensor library onto a legacy, non-AVX code path (a
CPU-capability override) makes the reference run, slowly but exactly, on
this machine --- so ground truth is available \emph{locally}, without a
second computer. We industrialized this into stage-validated porting: a
single script hooks all 31 stages of the reference tower (embeddings, 27
encoder layers, the window-attention merger, the final layer norm, the
downsampling projector) and dumps each output for a test image; the port
then proceeds stage by stage, and no stage is trusted until it matches its
reference dump to $\sim$$10^{-3}$ absolute (in practice every passing stage
matched to $10^{-4}$ or better).

\subsection{The encoder}
A SigLIP2 encoder layer --- layer norm, biased QKV projections,
bidirectional softmax attention (16 heads, dimension 72), residual, layer
norm, MLP with tanh-GELU, residual --- is a standard pre-norm vision
transformer block; notably, SigLIP2 has no rotary embedding (position
information enters once, additively, at the patch embedding), which makes
the block \emph{simpler} than its language-side counterparts. Validation:
patch embedding exact to $10^{-6}$ (it is one GEMM), and the first seven
encoder layers match the reference at $5.1\times10^{-5}$ max per element.

\subsection{Position embeddings, and the tie-breaking rule}
\label{sec:ties}
The one stage that failed was the one that looked most trivial. NaViT-style
resampling computes, for a $g_h\times g_w$ patch grid, fractional
coordinates $k/g$ and buckets them against 69 boundaries $j/70$ ---
integer index arithmetic, no learned weights involved. Our NumPy
reimplementation matched the reference on the $32\times32$ test grid, then
failed on a $28\times36$ grid --- by $1.8$ max per element, in an embedding
whose values span $\pm2$: one row of patches took its position embedding
from the wrong bucket row.

The root cause: at $k{=}10$, $g{=}28$, the fractional coordinate
$10/28 = 25/70$ is an \emph{exact rational tie} with boundary 25. In exact
arithmetic the reference's tie rule (``right'') puts it in bucket 25; but
both sides of the comparison are float32 values computed by different
arithmetic (a vectorized fused multiply-add inside the reference's range
constructor), and which side of the boundary each lands on is
\emph{implementation-defined rounding}, not semantics. We tried three
reimplementations --- float64 (then cast), float32 mimicry, and exact
integer arithmetic $\lfloor 70k/g\rfloor$ --- and each disagreed with the
reference on \emph{some} grid in $\{1,\dots,70\}$ (grids 25, 28, 49, 50,
and 55 flip, each differently, because the tie set depends on
$\gcd(g,70)$). There is no portable reimplementation of an
implementation-defined tie.

The fix is a rule we now apply generally: \textbf{for index arithmetic
with reachable exact ties, call the reference operator; do not reimplement
it.} The preprocessor imports the reference library (already present for
checkpoint reading) and evaluates exactly the two-line bucketization the
model uses --- small host-side tensor ops that are safe even on the
degraded-CPU path. After the change, positions match to $2\times10^{-6}$
on every grid. A secondary lesson for debuggers: our first attribution
split the error into a ``patch term'' and a ``position term'' that turned
out to be the same number algebraically --- the two diffs were identities
of each other, not independent evidence. Component attribution must
compare each component against \emph{its own} reference output (here,
hooking the reference's convolution output separately), not against
algebra that cancels.

\subsection{The window-attention merger, the only new kernel}
The mid-stack merger is genuinely new code: layer-norm the 1024-patch
sequence; reorder it so each $2\times2$ spatial window is contiguous (a
precomputed index gather); run self-attention \emph{within each 4-token
window} (a dedicated kernel: one block per (window, head), the $4\times4$
score matrix and its softmax entirely in shared memory); scatter back to
spatial order; add the residual; then concatenate each window's 4 vectors
(4608 wide), layer-norm, and project through the wide MLP with tanh-GELU,
adding the window-mean as a residual --- output one vector per window, 256
from 1024. The downsampling projector after the final layer is the same
merge shape with exact-GELU and \emph{no} mean-residual --- two
easy-to-miss asymmetries (activation flavor; residual presence) that the
stage gate catches immediately if transcribed wrongly. A further reuse
dividend: the window-index computation that orders patches for the
merger's attention is the \emph{same} grouping the concatenation step
needs, and the same formula again (on the halved grid) for the final
projector.

Validation: merger output $3.2\times10^{-4}$ max per element against its
reference dump; full tower --- patch embedding through projector, 64
language-space vectors --- $1.4\times10^{-5}$. The tower runs as a
resident GPU service holding the FP32 weights, beside the language engine.

\subsection{Variable resolution and multi-slice}
Large images arrive as several independent $\sim$$1000$-patch slices. The
reference encoder processes them packed in one sequence with
block-diagonal attention (explicit cumulative-length boundaries); reading
that code establishes that slices never attend across boundaries, so the
port may run the tower once per slice and concatenate outputs ---
verified end-to-end: a five-slice $896\times672$ test image matches the
packed reference at $1.2\times10^{-4}$ over all $5\times63$ output
vectors. The preprocessor reproduces the reference's slice packing
(patchify layout, per-slice grids) and emits, per slice, the patch matrix,
gathered position embeddings, and the two window-index vectors --- all
shape-driven, nothing learned.

\section{Two cuBLAS Attention Rewrites}
\label{sec:attn}

Sections~\ref{sec:cublas}--\ref{sec:chunked} drove projections and the
recurrence to the \textsc{cuBLAS} ceiling, but softmax attention remained
a hand-written kernel: one block per (position, head), a serial
dot-product loop over the history, a single-thread softmax reduction. Two
workloads exposed it from two directions --- the vision tower
(bidirectional attention over 1024 patches, 27 layers deep) and
long-context prefill (causal attention over a growing history) --- and
both were fixed by the same rewrite with the same memory trick.

\subsection{The layout trick}
Attention scores $S = QK^\top$ per head are a genuine matrix product, so
they belong in \textsc{sgemm}. The obstacle on a 6\,GB card is that $S$ is
large ($\text{heads}\times N\times P$ floats), and a GEMM normally wants a
fresh contiguous buffer. The observation that makes the rewrite free: the
engine's \emph{existing} per-(position, head) attention-score buffer,
row-major over positions and heads, is byte-for-byte a valid
\emph{column-major} score matrix for head $h$ if the GEMM is issued with a
leading dimension of $\text{heads}\times\text{seq}$ --- each score row
becomes column $b$ of $S_h$. The vendor GEMM writes scores \emph{into the
buffer the old kernel already owned} (zero new device memory), a batched
softmax kernel normalizes each contiguous column with a parallel tree
reduction, and a second GEMM multiplies by $V$. Grouped-query attention
falls out by pointing multiple query heads' GEMMs at the same key/value
sub-block.

\subsection{Tower attention: $6\times$}
In the tower (no causality, all positions valid) the rewrite is two GEMMs
and a plain column softmax per layer. Encoding a 1024-patch image drops
from 5.7\,s to \textbf{0.93\,s} ($6.1\times$); a five-slice image from
27.5 to 4.6\,s --- with outputs unchanged at $1.2\times10^{-4}$ vs.\ the
reference (the softmax reduction order changes, the results do not). The
naive kernel had been $\sim$85\% of tower time.

\subsection{Prefill attention: finding and removing the $O(N^2)$ wall}
\label{sec:wall}
On the language side the issue only appears at length. We built a tiered
prompt --- one document truncated to 2k, 5k, 8k, and 10k tokens --- and
measured prefill at each tier (Table~\ref{tab:tier}). The collapse is
unmistakable: 114\,\tps{} at 2k, 21 at 10k, and the wall-clock grows
\emph{faster} than quadratically (the kernel's serial history loop also
loses coalescing as the history grows). At 10k tokens, over 90\% of prefill
is one kernel.

The causal rewrite adds two details to the tower version: scores for
positions beyond each query's causal horizon are garbage from the full
rectangular GEMM, so the masked softmax normalizes only the valid prefix
$ps{+}b{+}1$ and \emph{zeroes the tail}, letting the $PV$ GEMM run the
full rectangle and multiply exact zeros --- trading a few wasted FLOPs
(cheap, Section~\ref{sec:cublas}) for kernel simplicity. Greedy outputs
are token-identical to the naive kernel on short prompts, and the planted
needle at 60\% depth of the 10k document --- a fabricated code phrase ---
is retrieved exactly both before and after the rewrite.

\begin{table}[t]
\centering
\caption{Tiered prefill on the production model, before and after
rewriting softmax attention as per-head \textsc{cuBLAS} GEMMs. The naive
kernel is invisible at 2k and is over 90\% of wall-clock at 10k; the GEMM path
is flat. Needle: a code phrase planted at 60\% depth of the full document,
retrieved exactly (only the 10k tier contains it).}
\label{tab:tier}
\begin{tabular}{lcccc}
\toprule
Prompt tokens & 2{,}000 & 5{,}000 & 8{,}000 & 10{,}000\\
\midrule
Naive attention (\tps{})  & 114.2 & 45.7 & 26.8 & 20.9\\
\quad prefill wall (s)    & 17.5  & 109.4 & 298.5 & 479.3\\
GEMM attention (\tps{})   & \textbf{408.5} & --- & --- & \textbf{361.2}\\
\quad prefill wall (s)    & \textbf{4.9} & --- & --- & \textbf{27.7}\\
\midrule
Speedup                   & $3.6\times$ & & & $\mathbf{17.3\times}$\\
Needle at 60\% depth      & \multicolumn{4}{c}{retrieved exactly (10k tier, before and after)}\\
\bottomrule
\end{tabular}
\end{table}

\begin{figure}[t]
\centering
\begin{tikzpicture}[font=\small]
  \def\xs{0.00042} 
  \def\ys{0.0095}  
  \draw[->] (0,0) -- (4.6,0);
  \node[below,font=\scriptsize] at (2.2,-0.42) {prompt tokens};
  \draw[->] (0,0) -- (0,4.2) node[left,font=\scriptsize]{prefill \tps};
  \foreach \x in {2000,5000,8000,10000} {
    \draw[gray!30] (\x*\xs,0) -- (\x*\xs,4.0);
    \node[below,font=\scriptsize,text=gray] at (\x*\xs,0) {\x};
  }
  \foreach \y in {100,200,300,400} {
    \draw[gray!20] (0,\y*\ys) -- (4.4,\y*\ys);
    \node[left,font=\scriptsize,text=gray] at (0,\y*\ys) {\y};
  }
  \draw[thick,red!70,mark=*,mark size=1.6pt]
    plot coordinates {(2000*\xs,114.2*\ys) (5000*\xs,45.7*\ys) (8000*\xs,26.8*\ys) (10000*\xs,20.9*\ys)};
  \node[right,font=\scriptsize,text=red!70] at (10000*\xs,20.9*\ys) {naive};
  \draw[thick,blue!70,mark=square*,mark size=1.6pt]
    plot coordinates {(2000*\xs,408.5*\ys) (10000*\xs,361.2*\ys)};
  \node[below right,font=\scriptsize,text=blue!70] at (10000*\xs,361.2*\ys) {GEMM attention};
\end{tikzpicture}
\caption{The $O(N^2)$ wall and its removal. The naive attention kernel's
throughput collapses with prompt length ($114\to21$\,\tps{}); the per-head
\textsc{cuBLAS} rewrite --- writing scores into the pre-existing attention
buffer, zero new device memory --- is nearly flat ($408\to361$\,\tps{}).
A 2k-token benchmark alone would never have shown the wall.}
\label{fig:wall}
\end{figure}
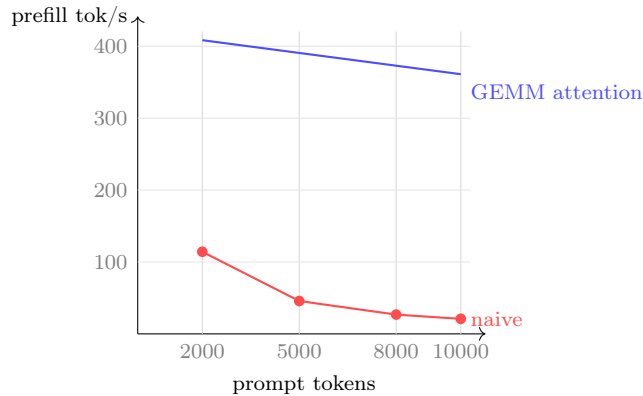

\section{Production Integration}
\label{sec:prod}

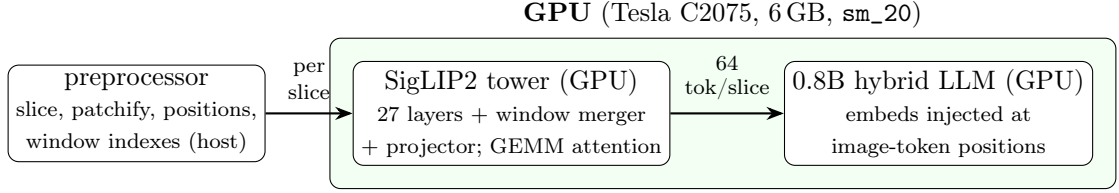
\begin{figure}[t]
\centering
\begin{tikzpicture}[font=\small,>=Stealth,
  box/.style={draw,rounded corners,align=center,minimum height=7.5mm,fill=white,inner sep=3pt},
  reg/.style={draw,rounded corners,inner sep=3.mm}]
  \node[box,minimum width=2.8cm] (prep) {preprocessor\\ \scriptsize slice, patchify, positions,\\ \scriptsize window indexes (host)};
  \node[box,minimum width=3.2cm,right=1.15cm of prep] (tower)
      {SigLIP2 tower (GPU)\\ \scriptsize 27 layers $+$ window merger\\ \scriptsize $+$ projector; GEMM attention};
  \node[box,minimum width=3.2cm,right=1.5cm of tower] (llm)
      {0.8B hybrid LLM (GPU)\\ \scriptsize embeds injected at\\ \scriptsize image-token positions};
  \draw[->,thick] (prep) -- node[above=0.5mm,font=\scriptsize]{\shortstack{per\\slice}} (tower);
  \draw[->,thick] (tower) -- node[above=0.5mm,font=\scriptsize]{\shortstack{64\\tok/slice}} (llm);
  \begin{scope}[on background layer]
    \node[reg,fill=green!5,fit=(tower)(llm),
      label={[font=\small]above:\textbf{GPU} (Tesla C2075, 6\,GB, \smtwenty)}] {};
  \end{scope}
\end{tikzpicture}
\caption{Production pipeline. The host preprocessor reproduces the
reference's slicing and index arithmetic (calling the reference operator
for bucketization, Section~\ref{sec:ties}); the resident tower encodes
each slice independently ($16\times$ compression); the language engine
replaces image-placeholder tokens with the tower's embeddings during
prefill.}
\label{fig:system}
\end{figure}

The serving layer mirrors the reference's prompt format exactly: each
image becomes an indexed tag, an overview segment of image-placeholder
tokens, and --- for sliced images --- per-row slice segments, every
delimiter a single token of the model's vocabulary. The engine replaces
each placeholder position with the next tower embedding during prefill.
Two correctness guards: image requests bypass the prefix-snapshot cache
(placeholder tokens are identical across images, so a cache hit would
silently reuse the \emph{previous} image's recurrent state --- with a
recurrent backbone this is a correctness rule, not a tuning choice), and
the tower's embedding count must equal the placeholder count by
construction (the preprocessor computes both from the same slice grids).

End-to-end production measurements, all on the serving path: a
single-image question (shape and color) answers in \textbf{1.7\,s}; an
image-plus-question prompt is $\approx$92 tokens ($16\times$ compression
leaves image cost almost invisible); a five-slice $896\times672$ image
with rendered text and shapes answers in $\approx$7\,s (measured on the
serving path; tower 4.6\,s of it), reading the rendered text exactly. An
eight-item smoke battery --- plain text, color/shape, rendered-text OCR,
two images in one prompt (correctly described \emph{per image}), the
multi-slice image, a multi-turn follow-up about an earlier image, plain
text after vision (cache-guard check), and Spanish --- passes in full.
Device memory at rest: 1.6\,GB engine (weights $+$ 16k-token KV cache $+$
score buffer) plus 2.7\,GB tower, 4.4 of 5.3 usable GB, leaving
$\approx$0.8\,GB headroom.

\section{Evaluation Summary}

On the same 2011 GPU as the companion study, the production system now
runs a modern multimodal assistant, entirely on-device. Language: decode
48.6\,\tps{} greedy; prefill flat at 361--408\,\tps{} through 10k tokens
($17\times$ at 10k over the naive attention path), 16k-token context
capacity,
exact needle retrieval at 60\% of 10k. Vision: the SigLIP2 tower matches
its reference to $1.4\times10^{-5}$ (multi-slice $1.2\times10^{-4}$),
encodes an image in 0.93\,s ($6\times$ over the naive kernel), and costs
64 language tokens per slice; a single-image question answers end-to-end
in 1.7\,s. Every stage of the vision port was gated on agreement with a
locally generated reference trace; the engine's greedy decoding is
token-identical to the reference implementation; and the two attention rewrites left
outputs unchanged (token-identical greedy decoding; tower agreement
unchanged).

The contrast with the companion 35B study~\cite{c2075moe} is the point.
There, the model did not fit, the design was forced into a streaming
GPU-prefill / integer-SIMD-CPU-decode hybrid, and the interesting results
were the systems work to make an over-sized model run. Here, the model
fits, the design collapses to a single all-GPU pipeline serving a
current-generation multimodal assistant, and the interesting results are a
prefill-throughput study, a stage-validated port, and a set of measured
reversals of intuition.

\section{Related Work}
The engine descends from \texttt{llama2.c}~\cite{llama2c} and its CUDA
port~\cite{llama2cu}, and shares its lineage and the larger MoE deployment
with the companion study~\cite{c2075moe}; low-bit weight handling follows
the post-training quantization literature~\cite{gptq,awq} and the integer
kernel style of \texttt{llama.cpp}~\cite{llamacpp}. The model is
MiniCPM-V-4.6~\cite{minicpmv,minicpmv46}; its vision encoder is
SigLIP2~\cite{siglip2} (sigmoid-loss pretraining~\cite{siglip}) with
NaViT-style variable-resolution position handling~\cite{navit}, and its
language backbone is a gated-delta-net hybrid~\cite{gateddeltanet} in the
Mamba lineage~\cite{mamba} from the Qwen3 family~\cite{qwen3}; the chunked
scan we implement is the standard intra-/inter-chunk formulation of that
family, here written by hand for \smtwenty{} and built from batched vendor
GEMMs. Our contribution is not any single technique but their all-GPU
composition on an unsupported architecture: the stage-validated port of a
modern vision stack, the tie-breaking rule for index arithmetic, the
zero-allocation GEMM formulation of both bidirectional and causal
attention, and the measured negatives (hand-GEMM ceiling, 4-bit decode)
that guided the design.

\section{Conclusion}
A fourteen-year-old GPU is now serving a current-generation multimodal
assistant, entirely on-device: a stage-validated hand port of a SigLIP2
tower with $16\times$ visual token compression, a hybrid delta-net
backbone running on dequantize-then-\textsc{sgemm} projections and a
chunked delta-rule scan, and a prompt path whose $O(N^2)$ attention wall
--- invisible at benchmark lengths, dominant at 10k tokens --- was removed
by two vendor-GEMM calls per head writing into memory the engine already
owned. The methodological refrain of this series held throughout, twice in
new forms: a benchmark is only as honest as its longest tier, and an index
computation with reachable exact ties cannot be reimplemented, only
invoked. On hardware the software ecosystem has abandoned, the performance
ceiling and the location of a bug are unreliable when reasoned about and
reliable when measured --- against a vendor BLAS call, a per-component
timer, a reference trace, a planted needle --- and that discipline, more
than any single kernel, is what turns a 2011 GPU into a working multimodal
inference engine.

\section*{Artifact and reproducibility}
The engine is a single CUDA/C source compiled with CUDA 8.0 for \smtwenty{}
(in a container) and linked against the CUDA 8.0 \textsc{cuBLAS}; the
vision tower is a second, self-contained CUDA source with the same build;
the preprocessor is a short host-side script that calls the reference
library only for image slicing and the bucketization operator. All
experiments run on one workstation; we report wall-clock seconds and
tokens/second directly, and GEMM efficiencies as a fraction of the
device's published FP32 peak. Stage-validation dumps are regenerable from
the public model checkpoint with the scripts described in
Section~\ref{sec:vision}.

\end{document}